\documentclass[twocolumn,showpacs,nofootinbib]{revtex4}
\usepackage{amssymb,epsfig,amscd}
\usepackage{graphicx}
\usepackage{dcolumn}
\usepackage{bm}

\usepackage{color} 
\usepackage[normalem]{ulem} 

\usepackage{epsfig}

\newcommand{\dirac}{\partial\llap{$\diagup$\kern-2pt}}

\def\be{\begin{equation}} 
\def\ee{\end{equation}}
\def\bq{\begin{eqnarray}} 
\def\eq{\end{eqnarray}}

\begin{document}
           

\title{Nucleation of quark matter in protoneutron star matter}

\author{B.~W.~Mintz$^a$, E.~S.~Fraga$^a$, G.~Pagliara$^{b,c}$, J.~Schaffner-Bielich$^{b,c}$}

\affiliation{$^a$ Instituto de F\'\i sica, Universidade Federal do Rio de Janeiro, 
Caixa Postal 68528, Rio de Janeiro, RJ 21945-970, Brazil\\
$^b$ Institut f\"{u}r Theoretische Physik, Ruprecht-Karls-Universit\"at,
Philosophenweg 16,  D-69120, Heidelberg, Germany\\
$^c$  ExtreMe Matter Institute EMMI, GSI Helmholtzzentrum f\"ur
Schwerionenforschung GmbH,
 Planckstra\ss e 1, D-64291 Darmstadt, Germany}

\begin{abstract}
The phase transition from hadronic to quark matter may take place already during the early 
post-bounce stage of core collapse supernovae when matter is still hot and lepton rich. If the 
phase transition is of first order and exhibits a barrier, the formation of the new phase occurs 
via the nucleation of droplets. We investigate the thermal nucleation of a quark phase in 
supernova matter and calculate its rate for a wide range of physical parameters. We show 
that the formation of the first droplet of a quark phase might be very fast and therefore the phase 
transition to quark matter could play an important role in the mechanism and dynamics of 
supernova explosions.
\end{abstract}

\pacs{PACS numbers: 26.50.+x, 64.60.Q-, 12.38.Mh, 25.75.Nq}

\maketitle

\section{Introduction}

The possibility that a first-order phase transition in dense matter
has implications for the explosions of supernovae was first proposed by
Migdal {\it et al.} about
30 years ago \cite{1979PhLB...83..158M}. 
Since then, 
a large number of papers addressed this issue with different hypotheses on the
nature of the phase transition: pion or kaon condensate matter or quark
matter, different models to compute the equation of state (EoS), different
simplifications for the complex hydrodynamical evolution of collapsing
massive stars and different approaches for the neutrino Boltzmann transport
\cite{1986Ap&SS.119...45T,1985PhRvL..55..126B,Gentile:1993ma}. However, 
only very recently was it possible to perform simulations using general
relativistic Boltzmann neutrino transport equations and adopting
realistic equations of state for quark matter \cite{Nakazato:2008su,Sagert:2008ka}. 
In Ref.~\cite{Sagert:2008ka}, in particular, it was shown that the phase
transition to quark matter can occur already during the early
post-bounce phase of a core collapse supernova event and that it produces a
second shock wave (the first being the usual shock wave after the 
bounce) which triggers a delayed supernova explosion, even within
spherical symmetry, for masses of the progenitor star up to 15
$M_{\odot}$. The formation of quark matter is also responsible for the
emission of a neutrino burst, typically a few hundred milliseconds after the
first neutronization burst, which could be detected by currently
available neutrino detectors, representing a spectacular possible signature of
quark matter formation in compact stars (see also \cite{Benvenuto:1989qr}). 

In the studies mentioned above, an important physical phenomenon is neglected 
for the sake of simplicity: the process of phase conversion in a first-order transition 
is actually driven by the nucleation of finite-size structures, such as droplets or 
bubbles, of the new phase within the old phase. The surface tension, $\sigma$, 
is the physical quantity that determines the nature of the process of phase conversion. 
If $\sigma$ is sufficiently 
small, nucleation can be very fast and the new phase is produced
almost in mechanical, thermal and chemical equilibrium with the nuclear matter phase. 
An intermediate value of $\sigma$ might render nucleation very difficult and the nuclear 
phase can be metastable for a significant amount of time. Then, the process of formation 
of the new phase, once triggered, would be a genuine nonequilibrium process, in which 
different mechanisms can take place: deflagration, detonation, convective instabilities 
(see, e.g., Refs. \cite{Drago:2005yj,Drago:2008tb} and references therein). Finally, nucleation 
is highly suppressed for large values of $\sigma$, and the formation of the new phase may only 
proceed via spinodal decomposition, if the density achieved is high enough to flatten out the 
activation barrier. (See Ref.~\cite{Bessa:2008nw} for a recent detailed discussion of 
the phase conversion process in a first-order phase transition.) 

The nucleation of quark matter in neutron stars has been explored mainly within a scenario, 
proposed in Ref.~\cite{Pons:2001ar}, in which the formation of quark matter occurs only when 
the protoneutron star is almost completely deleptonized and the temperature has already
dropped to, say, $1$ MeV. Under these conditions, quantum nucleation has been shown to 
be the most important mechanism for the formation of a quark phase
\cite{Iida:1997ay,Iida:1998pi,Berezhiani:2002ks,Bombaci:2004mt,Bombaci:2008wg}, 
also when color superconducting quark phases are present 
\cite{Drago:2004vu,Bombaci:2006cs}.
A less explored scenario is the nucleation of quark matter in hot and lepton-rich protoneutron 
stars. Refs.~\cite{Horvath:1992wq,Olesen:1993ek} present the first estimates and calculations 
showing thermal nucleation to be very efficient for temperatures of roughly $10$ MeV and 
practically negligible for temperatures below $2$ MeV. The simplifying assumption adopted 
in those papers (no leptons are included in the quark equation of state) might be, however, 
not realistic considering that for the large temperatures needed to nucleate quark matter 
the neutrino mean free path is small, and therefore neutrinos are trapped. In the presence of neutrinos, 
the critical densities for the phase transition are shifted toward larger values compared to a 
deleptonized neutron star. Moreover, as we will discuss in the following, the assumption 
of flavor conservation during the phase transition (also adopted in Ref.~\cite{Lugones:1997gg}) 
might be too conservative in light of the quark density fluctuations that are evidently present.

The main goal of this paper is to compare the time scale associated with the phase conversion 
with the dynamical time scale of core collapse supernovae during which quark matter might be 
eventually formed. In particular, since the explosion process occurs within a time scale of few 
hundred ms, the nucleation time must be of the same order of magnitude if quark matter plays 
indeed a role in the explosion mechanism. 

For this purpose, we reconsider thermal nucleation of quark matter within the scenario proposed 
in Ref.~\cite{Sagert:2008ka} of a phase transition occurring already during the early post-bounce 
stage of a core collapse supernova. This implies a value for the critical density $n_{c}\lesssim 2 n_0$ 
(where $n_0=0.16$ fm$^{-3}$ is the nuclear matter saturation density).  We systematically investigate 
the windows of free parameters of the model adopted to compute the equations of state and the 
corresponding nucleation time scale. Our strategy is to present {\it underestimates} of this 
time scale, so that a successful phase conversion could somehow constrain the equation of 
state parameter space. We also discuss the important issue of flavor number conservation during 
the phase transition. We argue that thermal nucleation might indeed be efficient under the conditions 
realized in a star soon after bounce. In such case, the phase transition would proceed very fast, thus confirming the results found in Ref.~\cite{Sagert:2008ka}. 

The paper is organized as follows. In Sec. II we present the model we adopt for the 
equation of state, as well as a brief self-contained description of homogeneous nucleation and
the role of statistical fluctuations. 
In Sec. III we present our results for the nucleation of nonstrange and strange quark matter. 
There, we also discuss quantum nucleation and the spinodal instability. 
Besides, we verify our equations of state by computing the stellar structure that emerge form the 
Tolman-Oppenheimer-Volkov (TOV) equations. Finally, Sec. IV contains our conclusions.

\section{Phenomenological framework}

\subsection{Equations of state}

The phase transition from nuclear to quark matter is implemented, as customary, 
by matching the equations of state for each phase. In order to do that, one has to 
choose appropriate models to compute the equation of state for each of 
the two phases and impose the conditions for mechanical, thermal and chemical
equilibrium to determine the transition point. For nuclear matter, we adopt the relativistic 
mean field model with the TM1 parametrization \cite{Shen:1998gq}, often used in 
supernovae simulations. For quark matter, we choose the MIT bag 
model \cite{Chodos:1974je,Farhi:1984qu} including perturbative QCD corrections 
\cite{Fraga:2001id,Alford:2004pf}.

Moreover, we consider two types of EoS for quark matter, both including the pressure 
from electrons and neutrinos, which are still present at this point of the stellar evolution. 
The first EoS contains only $up$ and $down$ quarks, while in the second we include a 
massive $s$ quark as well. 
The quark model has as free parameters the bag constant $B$, the 
mass of the strange quark $m_s$ (when present), and the value of the coefficient $c$, 
that accounts for the perturbative QCD corrections to the free gas pressure as follows: 
\begin{equation}\label{eq:pressure}
 p(\{\mu\}) = (1-c)\left[\sum_{i=u,d}\frac{\mu_i^4}{4\pi^2}\right] + p_s 
 + \frac{\mu_e^4}{12\pi^2}+\frac{\mu_\nu^4}{12\pi^2} - B
\end{equation}
where $B$ is the bag constant, $p_{s}$ is the contribution from the (massive) strange quark
\begin{equation}\label{eq:pressure_strange}
 p_s = (1-c) \frac{\mu_s^4}{4\pi^2} - \frac{3}{4\pi^2}m_s^2\mu_s^2,
\end{equation}
and terms ${\cal O}(m_s^4/\mu_s^4)\sim1\%$ in Eq. 
(\ref{eq:pressure_strange}) have been neglected. 
Notice that, for the u-d equation of state, we 
neglect the terms related to the {\it strange} quark.

The free parameters are fixed by requiring a critical density for the phase transition 
below two times the nuclear saturation density $n_0=0.16$ fm$^{-3}$ for the typical 
conditions of matter in the core of a star during a supernova collapse, i.e. temperatures 
$T=10-20$ MeV and lepton fractions $Y_L=0.3-0.4$. In this way, we fulfill our initial 
hypothesis of formation of quark matter in the early post-bounce stage. Another 
important criterion to fix our free parameters comes from the computation of the
maximum mass of cold hybrid stars. Taking into account the recent
measurement of the mass of PSR J1903+0327, $M=(1.671 \pm 0.008)M_{\odot}$ 
\cite{Freire:2009dr}, we require a maximum mass in agreement with this value. 
Finally, we investigate two possible scenarios for the appearance of strange quarks 
in the system. Since the nuclear EoS does not contain strangeness, a phase transition 
directly to strange quark matter might be difficult if we consider the slowness of weak 
reactions producing strange quarks with respect to the fast deconfinement/chiral
phase transition process driven by the strong interaction. Therefore, we discuss a 
first case in which the phase transition involves two-flavor quark matter (strange quarks 
will be produced only later, via weak interaction, as suggested in Ref. \cite{Bhattacharyya:2006vy}). 
In the second scenario, we consider a fast production of strange quarks:
since we assume critical densities of the order of 2 times the saturation density and 
temperatures of a few tens of MeV, it is possible that a small seed of strange matter appears 
in the system through hyperons or kaons \cite{Norsen:2002qw}. Once strangeness is produced 
in the hadronic matter, this would trigger the phase transition directly to strange quark matter.

The equations of state for nuclear matter and quark matter are calculated under conditions 
of local charge neutrality, local lepton fraction conservation (i.e., the two phases have the 
same $Y_L$), and weak equilibrium. Under these assumptions, the conditions of phase 
coexistence, as found in Ref. \cite{Hempel:2009vp}, are the equality of the total pressure of 
the two phases $P^H = P^Q$ and the following condition of chemical equilibrium:
\begin{equation}\label{eq:chem_equil}
 \mu_n+Y_L \mu_{\nu}^H=\mu_u+2\mu_d+Y_L\mu_{\nu}^Q \equiv \mu_{eff}, 
\end{equation}
where $\mu_n$ and $\mu_{\nu}^H$ are the chemical potentials of neutron and neutrinos 
within the nuclear phase, and $\mu_u$, $\mu_d$ and $\mu_{\nu}^Q$ are the chemical 
potentials of up and down quarks and of neutrinos within the quark phase, respectively. 
The quantity $\mu_{eff}$ is an effective chemical potential, which is always 
the same in both phases. Notice that the condition (\ref{eq:chem_equil}) is always valid,
although the condition $P^H=P^Q$ is valid only in the transition point.
Here we use the zero-temperature equations of state, since a temperature of the order of 
a few tens of MeV does not alter considerably the equation of state \footnote{For a free 
massless gas, the corrections would be ${\cal O}(T^2/\mu^2)\sim 1\%$.}.
 
Moreover, we assume that the different degrees of freedom in both 
phases are in chemical equilibrium with respect to weak reactions (see
Sec. II C):
\begin{eqnarray}
 \mu_n+\mu_{\nu}^H &=& \mu_p+\mu_e^H \\
 \mu_d+\mu_{\nu}^Q &=& \mu_{u}+\mu_e^Q \\
 \mu_d &=& \mu_s
\end{eqnarray}

Finally, the conditions of local charge neutrality and local lepton fraction 
within the two phases allow us to compute all chemical potentials 
in terms of only one independent chemical potential:
\begin{eqnarray}
 n_p &=& n_e^H\\
 \frac{2}{3}n_u-\frac{1}{3}n_d-\frac{1}{3}n_s &=& n_e^Q\\
\label{eq:electric_neutrality}
 \frac{n_e^H+n_{\nu}^H}{n_B^H} &=& \frac{n_e^Q+n_{\nu}^Q}{n_B^Q} = Y_L,
\end{eqnarray}
where $n_i$ ($i=n,p,u,d,s,e,\nu$) are the densities of the different 
species of particles. 

Notice that fixing $Y_L$ locally results in a jump of the chemical potential 
of neutrinos at the interface of the phase transition. This might be not very 
realistic, as discussed in Ref. \cite{Pagliara:2009dg} where it has been shown 
that a mixed phase should instead be considered due to the global conservation 
of the lepton number. We leave the discussion of nucleation in mixed phase for a 
future study. 

\subsection{Thermal homogeneous nucleation}

First-order phase transitions are very well known even from simple everyday examples, 
such as in the melting of ice. In such a case, the conversion from one phase to the other 
usually occurs slowly and very close to the thermodynamical equilibrium, following 
the so-called Maxwell construction. However, when some relevant 
external control parameter (such as the temperature or the density) 
changes abruptly when a system is near the transition, the system finds itself in an 
unstable situation. For definiteness, consider a system initially homogeneous 
in a low-density phase (a ``gas''), and close to the transition 
line to a high-density (``liquid'') phase. Now, let it suffer a sudden 
compression. Although the system was prepared at the 
gas phase, the free energy at the new, higher density disfavors the gas phase 
and the liquid phase now becomes the stable one: phase conversion is about to begin. 

The ever-present thermal and quantum fluctuations will not be suppressed, 
as expected in equilibrium, due to the instability of the system. Such 
fluctuations will drive the system to another point of stability 
of the phase diagram. The evolution in time of those fluctuations 
are at the heart of our discussion. 

For first-order phase transitions, there can be two kinds of instabilities 
that dominate the dynamics of the phase conversion \cite{Gunton_et_al}. If 
a homogeneous system is brought into instability close enough to the 
coexistence line of the phase diagram, its dynamics will be dominated by 
large-amplitude, small-ranged fluctuations. In these cases, large 
amplitudes are necessary for the development of the phase transition 
once the system is in a metastable equilibrium. Those are usually 
referred to as bubbles (or droplets) and the process that 
creates them is called nucleation. In the other case, if the external 
perturbation is big enough and the system finds itself far from the coexistence 
line, the dominant fluctuations will have small amplitudes and large wavelengths, 
the process of phase conversion is called spinodal decomposition. 
In this work, we focus on thermal nucleation of quark matter as nuclear 
matter is compressed in a stellar collapse, leaving a discussion on a possible role 
for spinodal decomposition and quantum nucleation to the final section. 

A standard, field-theoretical approach
for thermal nucleation in one-component metastable systems 
was developed by Langer in the late sixties \cite{Langer:1969bc}. In this formalism, 
a key quantity for the calculation of the rate of nucleation is the coarse-grained free 
energy functional
\begin{equation}\label{eq:coarse-grainedFunctional}
 F[\phi] = \int d^3r\,\left\{\frac12[\nabla \phi({\bf r})]^2 + V[\phi({\bf r})]\right\},
\end{equation}
where $\phi({\bf r})$ is the order parameter of the phase transition at a 
given point ${\bf r}$ of space. By 
assumption, the ``potential'' $V(\phi)$ has a global (true) minimum at $\phi_t$ 
and a local (false) one at $\phi_f$. 
At a given baryon chemical potential $\mu$ of the metastable phase,
the difference $\Delta V\equiv V(\phi_t) - V(\phi_f)$ 
is identified with the pressure difference between the stable and the metastable phases, 
with opposite sign: $\Delta V = - \Delta p(\mu) = p_t-p_f$, where $p_t$ ($p_f$) is the 
pressure for the true (false) phase at baryon chemical potential $\mu$.

The field equation for $\phi({\bf r})$ is given by a minimum of the 
functional $F$. One can easily think of three static solutions. Two 
of them are the trivial ones given by homogeneous field configurations 
with $\phi({\bf r})=\phi_t$ or 
$\phi({\bf r})=\phi_f$. The third is a spherically symmetric 
bubblelike solution that has as boundary conditions 
\begin{eqnarray}\label{eq:bubbleboundaryconditions}
&& \phi(r=0) = \phi_t , \crcr 
&& \phi(r\rightarrow\infty) = \phi_f.
\end{eqnarray}
Roughly speaking, this means that the stable phase is found deep in the bubble 
and the metastable one is found away from it. Somewhere in-between, the order 
parameter must change from its central value $\phi_t$ to $\phi_f$ at 
$r\rightarrow\infty$. The relatively thin region which marks the border 
between ``inside'' ($\phi=\phi_t$) and ``outside'' ($\phi=\phi_f$) 
the bubble is called the bubble wall. 

Exactly at the coexistence line, one can prepare one (infinite) system 
with the two homogeneous phases in equal proportions divided by 
a plane wall with a small width. This configuration is static, once 
no phase is favored. Further, each phase occupies a semi-infinite 
volume. If the system is slightly pushed into metastability, the static 
solution for $\phi({\bf r})$ is a bubble with a very large radius and 
still a small wall width. This is the starting point for the {\it thin-wall 
approximation}: the free energy (\ref{eq:coarse-grainedFunctional}) of 
the system of volume $(4\pi/3)L^3$ $(L\rightarrow\infty)$ is 
determined by the outcome of 
a competition between a surface energy term, which is positive and comes 
from $|\nabla\phi|^2$ in (\ref{eq:coarse-grainedFunctional}), and a bulk term, 
which is negative and corresponds to the potential $V$, or to the pressure 
difference between the phases. 
Notice that, within this approximation, $\phi(r)$ is constant, except over the (thin) wall 
of the bubble, and so $V(\phi)$ is also essentially constant both inside and 
outside the bubble. This means that the free energy for the bubble 
configuration of radius $R$ in the thin-wall approximation of Eq. 
(\ref{eq:coarse-grainedFunctional}) is given by
\begin{equation}\label{eq:F_bubble}
 F_{\rm bubble}(R) = 4\pi R^2\sigma - \frac{4\pi}{3}(L^3-R^3)p_f - \frac{4\pi}{3}R^3p_t
\end{equation}
whereas the homogeneous metastable configuration has $|\nabla \phi|^2=0$ and 
\begin{equation}\label{eq:F_metastable}
 F_{\rm metastable} = - \frac{4\pi}{3}L^3p_f,
\end{equation}
In Eq. (\ref{eq:F_bubble}), we introduced the surface tension $\sigma$, which is merely 
the energy per unit area of the bubble 
wall. 
As will be clear below, it is a key physical 
quantity in our analysis.

According to the standard theory \cite{Langer:1969bc}, the nucleation rate has as its 
main ingredient the free energy shift when a bubble is created from fluctuations in the 
homogeneous metastable phase. From to Eqs. (\ref{eq:F_bubble}) and (\ref{eq:F_metastable}) 
we have
\begin{eqnarray}\label{eq:DeltaF}
 \Delta F(R)\equiv &&\!\!\!\!\! F_{\rm bubble}(R) - F_{\rm metastable} \crcr
             =&&\!\!\!\!\! 4\pi R^2\sigma - \frac{4\pi}{3}R^3(\Delta p),
\end{eqnarray}
where $\Delta p = p_t - p_f > 0$. Here, the pressures in each of the phases
are calculated for the same value of $\mu_{eff}$. Notice that 
this implies different baryon chemical potentials and densities for each phase, due to the 
conditions (\ref{eq:chem_equil})-(\ref{eq:electric_neutrality}).

Bubble configurations of given radii $R$ arise from the homogeneous 
metastable phase due to thermal fluctuations, and each of 
those has an associated value of $\Delta F(R)$. From Eq. (\ref{eq:DeltaF}), 
we can see that $\Delta F(R)$ has a maximum at the critical radius 
$R_c \equiv 2\sigma/\Delta p$. The equations of motion show that 
any bubble with $R<R_c$ will shrink and disappear whereas any bubble 
with $R>R_c$ will grow, as a consequence of the competition between the positive 
surface energy and the negative bulk energy. Hence, the critical bubbles 
are the smallest bubbles that can start to drive the phase conversion dynamics. 
To give a quantitative meaning to the process of nucleation, 
one can calculate the rate $\Gamma$ of critical bubbles created by 
fluctuations per unit volume, per unit time. In Langer's formalism \cite{Langer:1969bc}: 
\begin{equation}\label{eq:Gamma_def}
 \Gamma = \frac{{\cal P}_0}{2\pi} \,\exp\left[-\frac{\Delta F(R_c)}{T}\right],
\end{equation}
where the prefactor ${\cal P}_0$ is usually factorized into two parts: a 
statistical prefactor, which measures the rate of successful creation of 
a critical bubble by thermal fluctuations, and a dynamical prefactor, which 
measures the early growth rate of the bubble. As customary, we adopt 
${\cal P}_0/2\pi=T^4$, which corresponds to an overestimate of the actual 
prefactor. (For an exact calculation of ${\cal P}_0$ see, e.g., Ref. \cite{Csernai:1992tj}.) 
This constitutes one of our main reasons to interpret our results as providing an 
overestimate for the nucleation rate. It goes in line with the thin-wall approximation, 
which is also known to overestimate $\Gamma$ when compared to the exact (numerical) 
result \cite{Scavenius:2000bb}. Although this overestimate can lead to an overall factor of 
$\sim 10^2$ or even higher \cite{Csernai:1992tj}, the qualitative aspects of the 
results shown in the next section can be barely changed. And, since we are concerned with 
providing {\it underestimates} for thermal nucleation time scales under core collapse 
supernovae typical conditions, these details are not relevant.

Our final formula for the nucleation rate reads 
\begin{equation}\label{eq:Gamma_final}
 \Gamma = T^4 \,\exp\left[-\frac{16\pi}{3}\frac{\sigma^3}{(\Delta p)^2 T}\right],
\end{equation}
where we used Eqs. (\ref{eq:DeltaF}) and $R_c=2\sigma/\Delta p$. Notice that 
the influence of the equation of state is present through $\Delta p$. Also, there is 
a remarkably strong dependence of $\Gamma$ on the surface tension $\sigma$, 
which will be determinant for the nucleation time scale. 

It is convenient to introduce the {\it nucleation time} $\tau$, defined as 
the time it takes for the nucleation of one single critical bubble inside a volume 
of $1km^3$ inside the core of the protoneutron star:
\begin{equation}\label{eq:tau}
 \tau \equiv \left(\frac{1}{1km^3}\right)\frac{1}{\Gamma}.
\end{equation}
This is the time scale to be compared with the duration of the early post-bounce phase
of a supernova event, few hundreds of milliseconds, during which
it has been shown that quark matter formation could trigger the explosion \cite{Sagert:2008ka}. 
With this definition, we assume that temperature and density are constant 
within this central volume of $1km^3$. Of course, this also goes in the direction 
of underestimating this time scale. In a more realistic calculation, one should first 
compute the pressure and density profiles using the TOV 
equations, then calculate the {\it local} value of $\Gamma$, and finally integrate over 
the region containing metastable matter. However, the density profiles are almost flat 
within the central kilometers of the star, thus making our assumption quite reasonable.

Finally, we note that we calculate the time of production of one single critical bubble, 
which has a typical size of some fermi. We do not study, in this paper, the growth regime 
of the {\it quark front}. Once again, this leads to an underestimate: in comparing $\tau$ with 
the bounce time scale as a criterion for the formation of a quark core, we tacitly assume 
that the quark matter bubble becomes macroscopic almost instantaneously, an obviously 
artificial simplification.

\subsection{The role of statistical fluctuations}

As discussed in the Introduction, a possible alternative approach to model the phase 
transition to quark matter assumes isospin (and strangeness) conservation during the 
phase transition \cite{Olesen:1993ek,Lugones:1997gg,Bombaci:2004mt}. This implies a
transition from hadronic matter in chemical equilibrium to an intermediate quark matter 
phase $Q^*$ in which quarks are not in chemical equilibrium. The chemical potentials 
of quarks are calculated, indeed, by assuming that the fractions of different quark flavors 
are the same, both in the hadronic and in the quark phase. This assumption is based on 
the argument that the time scale of the phase transition is regulated by QCD, thus typically 
of the order of $10^{-23}~$s, and much faster than the weak interaction time scale (weak
interactions would produce quark matter in chemical equilibrium $Q^{eq}$ only after the 
phase transition is completed). 

However, as noticed in Refs. \cite{Norsen:2002qw,DiToro:2006pq}, statistical fluctuations 
of the number densities of quarks can have important effects on nucleation. To estimate these 
effects one can calculate the phase transition point by considering the two equations of state 
in chemical equilibrium. At the critical density one compares the average numbers 
$N^{eq}_{u,d}$ of up and down quarks within $Q^{eq}$ and within the nuclear phase $N^{*}_{u,d}$ 
in a fixed volume $V$.  If these numbers are different only by $\sim 3\sqrt{N^*_{u,d}}$, it means 
that most probably fluctuations would drive the transition directly to $Q^{eq}$.

The volume $V$ in which we consider fluctuations corresponds to the volume of a drop of the 
new phase with critical radius $R_{crit} = 2\sigma/\Delta P$.  Again, the surface tension $\sigma$ 
is the crucial quantity that determines whether the phase transition occurs via the intermediate 
phase $Q^*$ or directly to $Q^{eq}$. Taking into account the uncertainties on the value of $\sigma$ 
we estimate the critical radii to be of order $\sim 6~$fm. Using the simple calculation explained 
above, we obtain that statistical fluctuations are efficient for radii of the order of $2-4~$fm and thus 
of the same order of magnitude as the critical radii. So, we assume, as in 
Ref. \cite{Berezhiani:2002ks}, that the first drop is nucleated already in the  $Q^{eq}$ phase.

Now that we have all the ingredients for the calculation of the nucleation 
times, we can proceed to our results in the next section.

\section{Results and discussions}

In order to evaluate if the time scale $\tau$ for the nucleation of quark matter 
is compatible with the bounce time scale $\tau_B$, we underestimate the time 
$\tau$ for the formation of a critical bubble as a function of density under various 
conditions of temperature, 
as well as for different equations of state and values 
of surface tension, given the scenario 
described in the previous section. 

Recent supernova simulations \cite{Sagert:2008ka} indicate that the 
central density of a protoneutron star can be as high as $2n_0$ during 
the bounce, and this value will serve as a cutoff density in our analysis. 
Still in the spirit of underestimating $\tau$ (or, equivalently, 
overestimating the nucleation rate $\Gamma$), we consider that nucleation 
is effective if $\tau<\tau_B\lesssim100$ ms for some $n<2n_0$. 

\subsection{Nucleation times for nonstrange matter}

As our first case, we consider the transition from beta-stable nuclear 
matter to beta-stable quark matter composed of $u$ and $d$ quarks, plus 
electrons and electron neutrinos, with a fixed lepton fraction $Y_L$. 
Later on, we will discuss the case of a transition from nuclear matter 
to u-d-s quark matter (both lepton-rich and in beta equilibrium). 

We start our analysis by the case of a low-density transition: $n_c=1.5n_0$. 
We assume a lepton fraction $Y_L=0.4$ and consider two values for the temperature, 
representing a ``minimum'' and a ``maximum'' value that can be expected during the bounce, 
and two values for the surface tension.

In Fig. 1 
one can see the behavior of the nucleation time of a single critical bubble (as defined 
in the previous section) versus the density, in units of $n_0$. 
\begin{figure}[!hbt]
\begin{center}
\vskip 0.2 cm
\includegraphics[width=8cm]{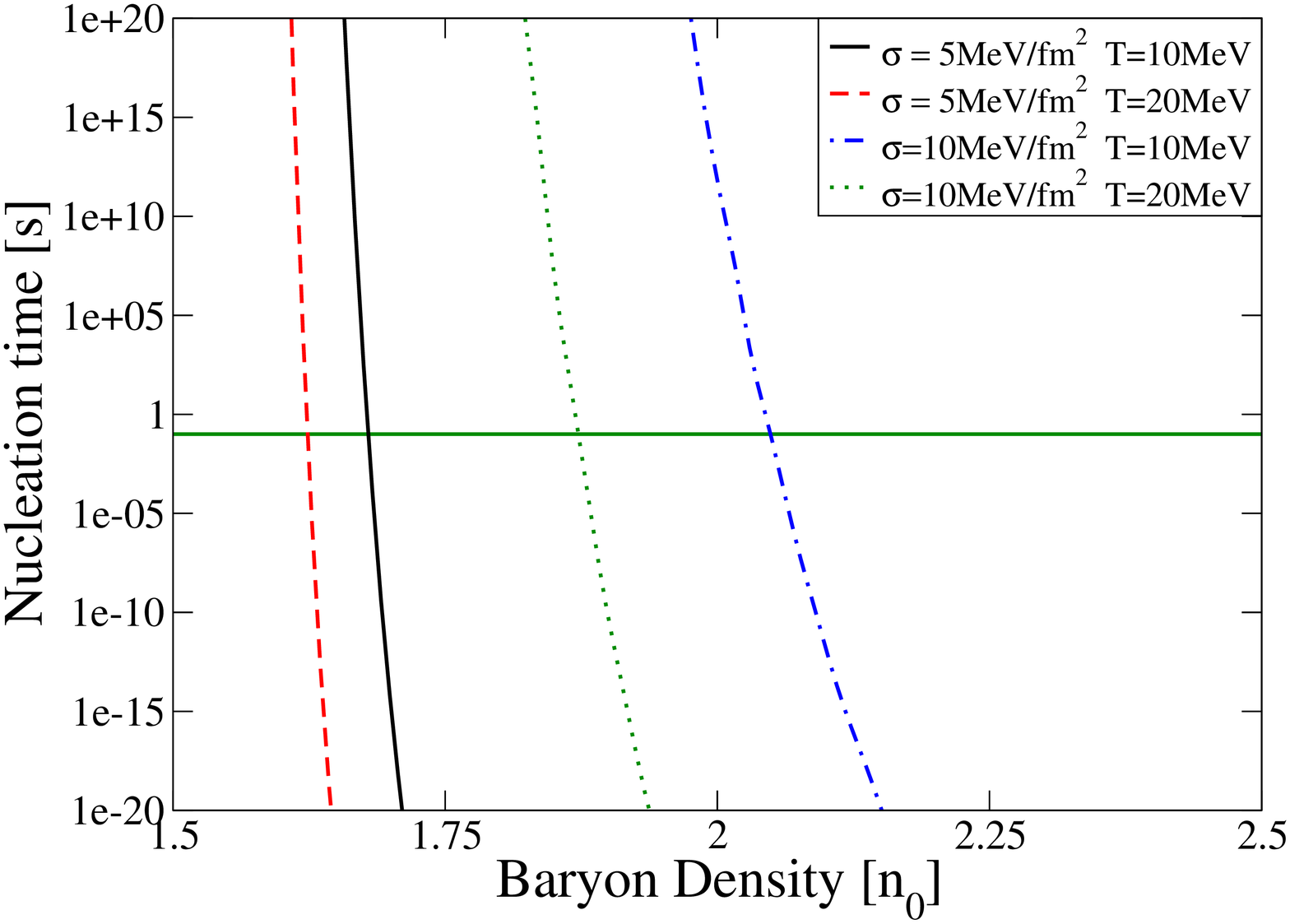}
\caption{Nucleation time as a function of baryon density for u-d 
quark matter ($n_c=1.5n_0$). The horizontal line corresponds to $\tau=100$ms.}
\end{center}
\label{fig:20n0YL04}
\end{figure}

\begin{figure}[!hbt]
\begin{center}
\vskip 0.2 cm
\includegraphics[width=8cm,angle=0]{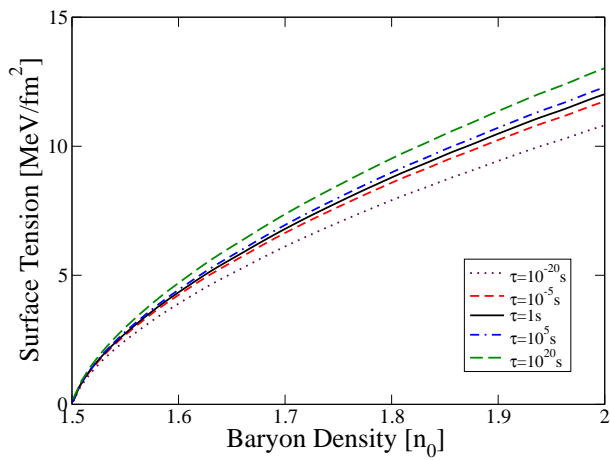}
\caption{Contour lines of constant nucleation time (contour lines) as a function of density 
and surface tension for u-d quark matter. Here, $n_c=1.5n_0$, $Y_L=0.4$, 
and $T=20~$MeV.}
\end{center}
\label{fig:12n0YL04}
\end{figure}

As expected, the nucleation time $\tau$ has an extremely strong dependence 
on both density (notice the logarithmic scale for $\tau$) and on the surface tension, a 
feature that can also be seen in Fig. 2. For low values of $\sigma$ nucleation 
becomes feasible at relatively low densities, although such densities increase 
steadily as the surface tension rises. However, if the (basically unknown) 
surface tension is larger, the density for $\tau\sim100$ ms may be higher 
than our $2n_0$ cutoff, and nucleation should not be an efficient mechanism 
for phase conversion. In this sense, we can 
expect that if the nuclear-quark matter transition occurs in this scenario 
the surface tension must be quite small.

We can also investigate how the choice of the critical density can affect the 
nucleation time (Fig. 3). In any case, a surface tension larger than 
roughly $15$ MeV/fm$^2$ 
seems to be sufficient to prevent thermal nucleation (for $T=20$ MeV or lower).

\begin{figure}[!hbt]
\begin{center}
\vskip 0.2 cm
\includegraphics[width=8cm,angle=0]{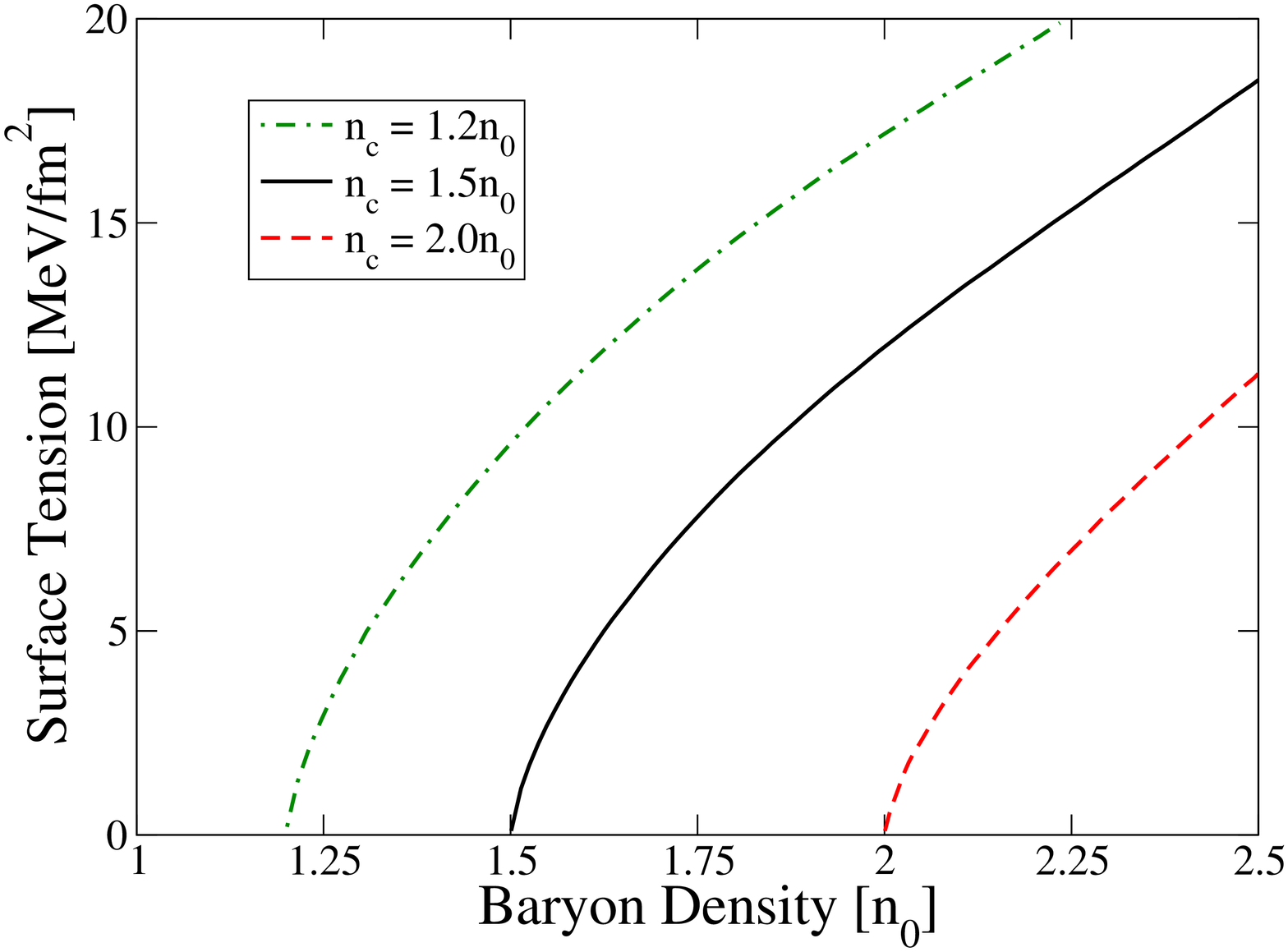}
\caption{Lines of constant nucleation time ($\tau=100~$ms) for 
$n_c/n_0=1.2,\,1.5\,,2.0$ with $T=20~$MeV, $c=0$ and $Y_L=0.4$.}
\end{center}
\label{fig:12n0_and_20n0YL04}
\end{figure}

In Fig. 4, we show the role played by the temperature in the process 
of nucleation. We may notice that the precise value of the temperature does not 
affect the nucleation times as strongly as the surface tension does, as long as 
it is kept in the range expected during the early post-bounce phase at the 
protoneutron star core, i.e., roughly from $10$ to $25$ MeV.
%
\begin{figure}[!hbt]
\begin{center}
\vskip 0.2 cm
\includegraphics[width=8cm,angle=0]{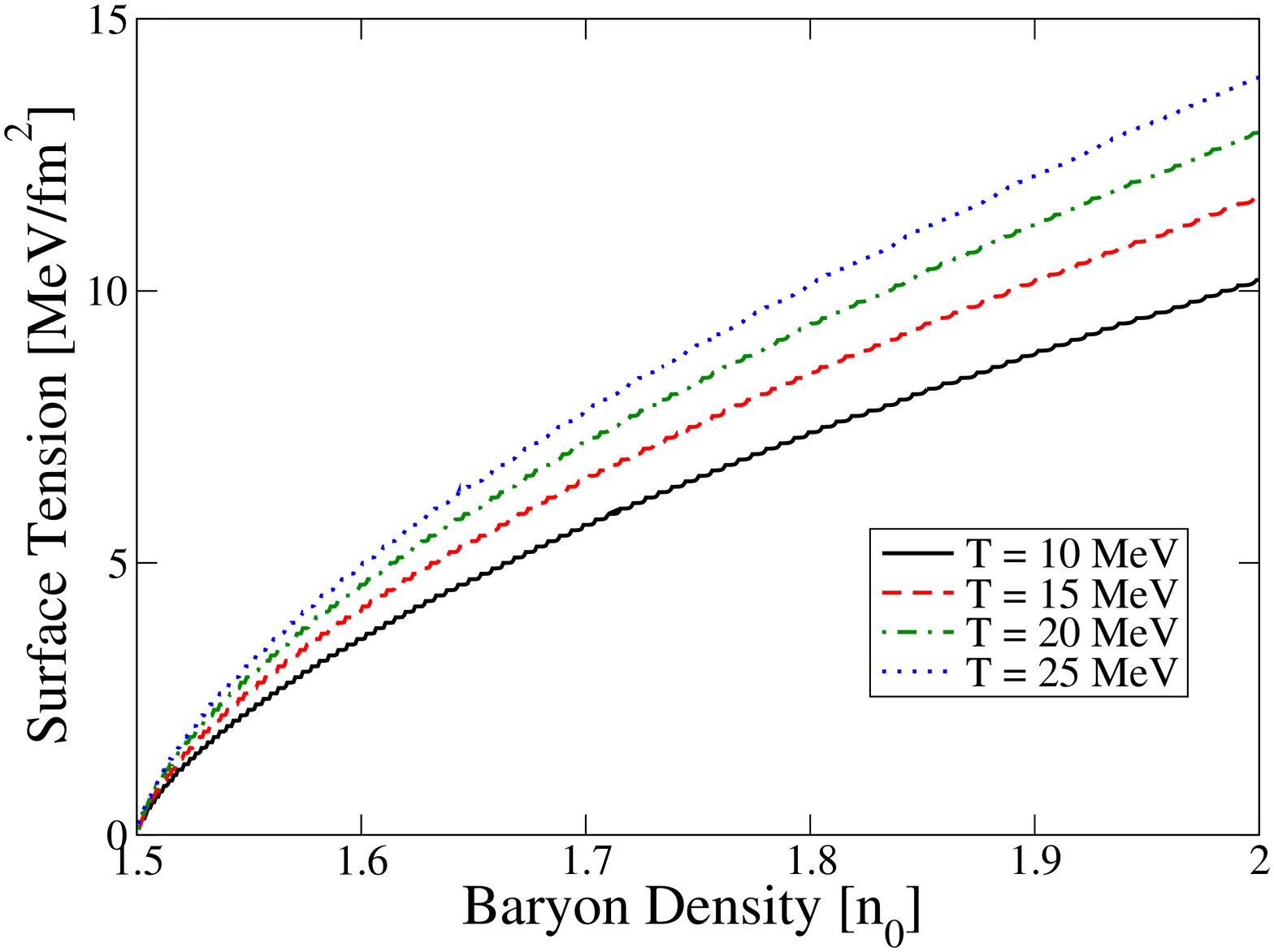}
\caption{Lines of constant nucleation time ($\tau=100$ms) 
for noninteracting u-d quarks, for temperatures between $10$ and $25$ MeV 
($n_c=1.5n_0$, $Y_L=0.4$). }
\end{center}
\label{fig:c02_vs_c0}
\end{figure}

Although the exact numbers should not be taken at face value, given the 
uncertainties involved, we believe that the order of magnitude of 
the actual limiting value of the surface tension for $\tau=100$ ms is correct.

\subsection{Nucleation of strange matter}

Although the core of a supernova progenitor star before its collapse does not 
contain any strangeness, the energy density achieved during and right 
after bounce allows for the presence of a small amount of hyperons in the hadronic 
phase \cite{Ishizuka:2008gr}. Such particles do not contribute significantly to the pressure 
or to the energy density, but density fluctuations of such hadrons may induce 
the formation of bubbles of strange quark matter.

The introduction of strange quarks makes the EoS stiffer, i.e. for a given baryon 
chemical potential $\mu$ the corresponding pressure becomes higher. Once the 
nuclear EoS is the same, $\Delta p$ will be higher for a given value of $\mu$, 
and therefore the nucleation rate will also be higher. Figure 5 shows 
a comparison between the transition from nuclear matter to either u-d or 
u-d-s quark matter for two values of the lepton fraction $Y_L$. 
We can notice that a decrease in $Y_L$ increases the efficiency of 
thermal nucleation. This is expected because deleptonization not only 
renders nuclear matter less stable but it also heats up the stellar 
core (the former effect, however, is not accounted for in this work). 

\begin{figure}[!hbt]
\begin{center}
\vskip 0.2 cm
\includegraphics[width=8cm,angle=0]{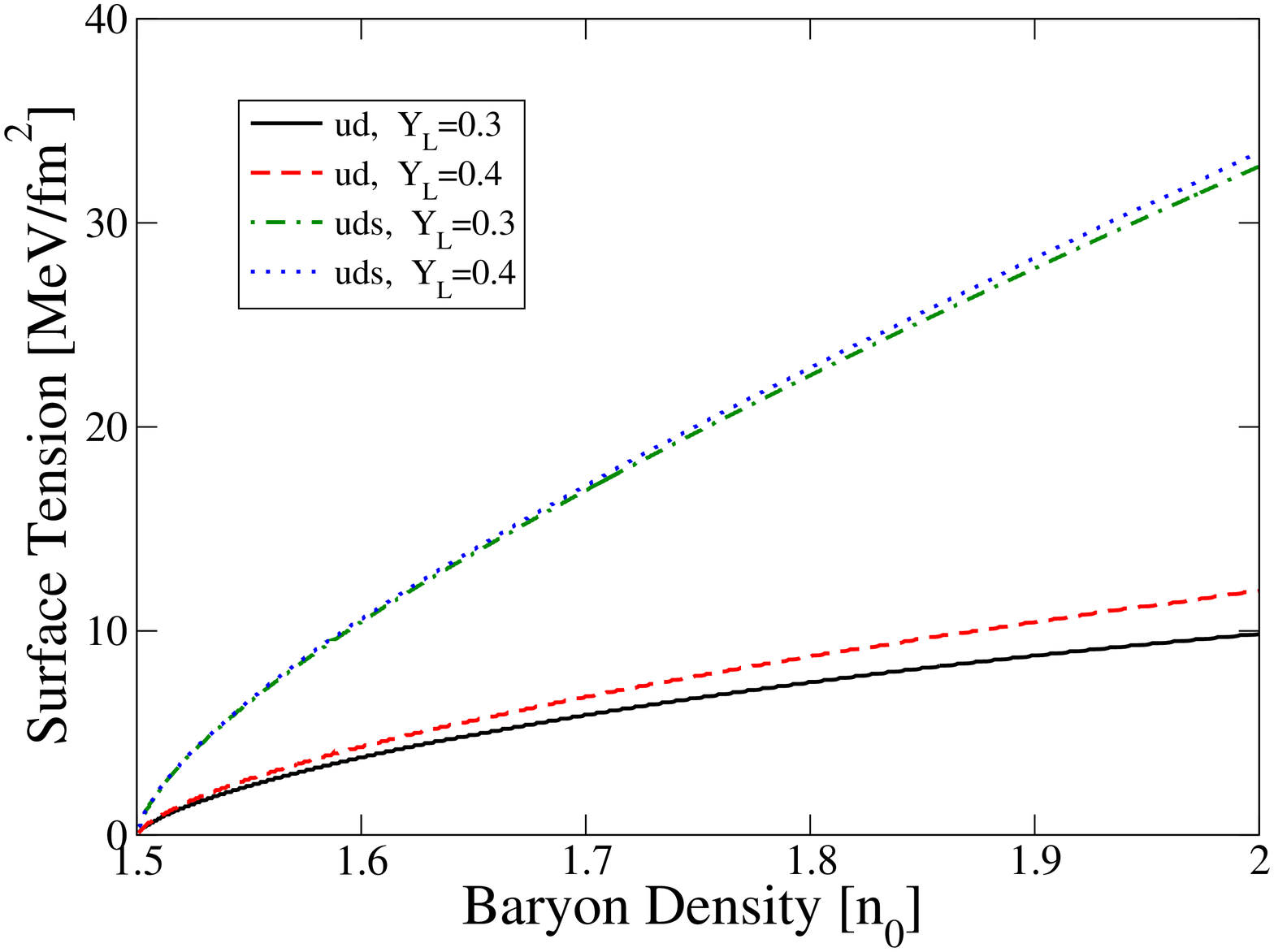}
\caption{Lines of constant nucleation time ($\tau=100~$ms) for 
the transition from nuclear matter to u-d or u-d-s quark matter, 
with $Y_L=0.3,\,0.4$ ($n_c=1.5n_0$, $T=20~$MeV, $c=0$ and $m_s=0$).}
\end{center}
\end{figure}

Up to now, we have only considered noninteracting quarks. Results 
from two-loop perturbative three-flavor QCD at finite density \cite{Fraga:2001id} 
show that strong interactions can be effectively accounted for in the 
equation of state by introducing a factor $c<1$ \cite{Alford:2004pf}, as in 
Eq. \ref{eq:pressure}. 
This factor makes the quark EoS softer in the pressure-chemical potential plane 
[see Eq. (\ref{eq:pressure})] and, therefore, $\Delta p$ should be smaller, making the 
nucleation time $\tau$ larger for a given density, according to Eqs.  
(\ref{eq:Gamma_final}) and (\ref{eq:tau}). 
As an explicit example, 
we compare the $c=0$ case with $c=0.3$, in the case of 
$n_c=1.5n_0$, as displayed in Fig. 6, where we also show the influence of 
strange quark mass $m_s$. Notice that the introduction 
of interactions via the parameter $c$ drastically increases the 
nucleation time, so that only for a low value of the surface tension,
e.g. $\sigma \sim 10$ MeV/fm$^2$ (for $Y_L=0.4$), nucleation can be efficient 
if the density reaches a value close to $2n_0$. \footnote{Of course, the surface 
tension should also be affected by loop corrections, and that could eventually 
reduce or even balance out this effect on the final nucleation dynamics. As 
becomes clear from all this analysis, a reliable estimate of the surface tension 
for cold dense matter is called for.}

\begin{figure}[!hbt]
\begin{center}
\vskip 0.2 cm
\includegraphics[width=8cm,angle=0]{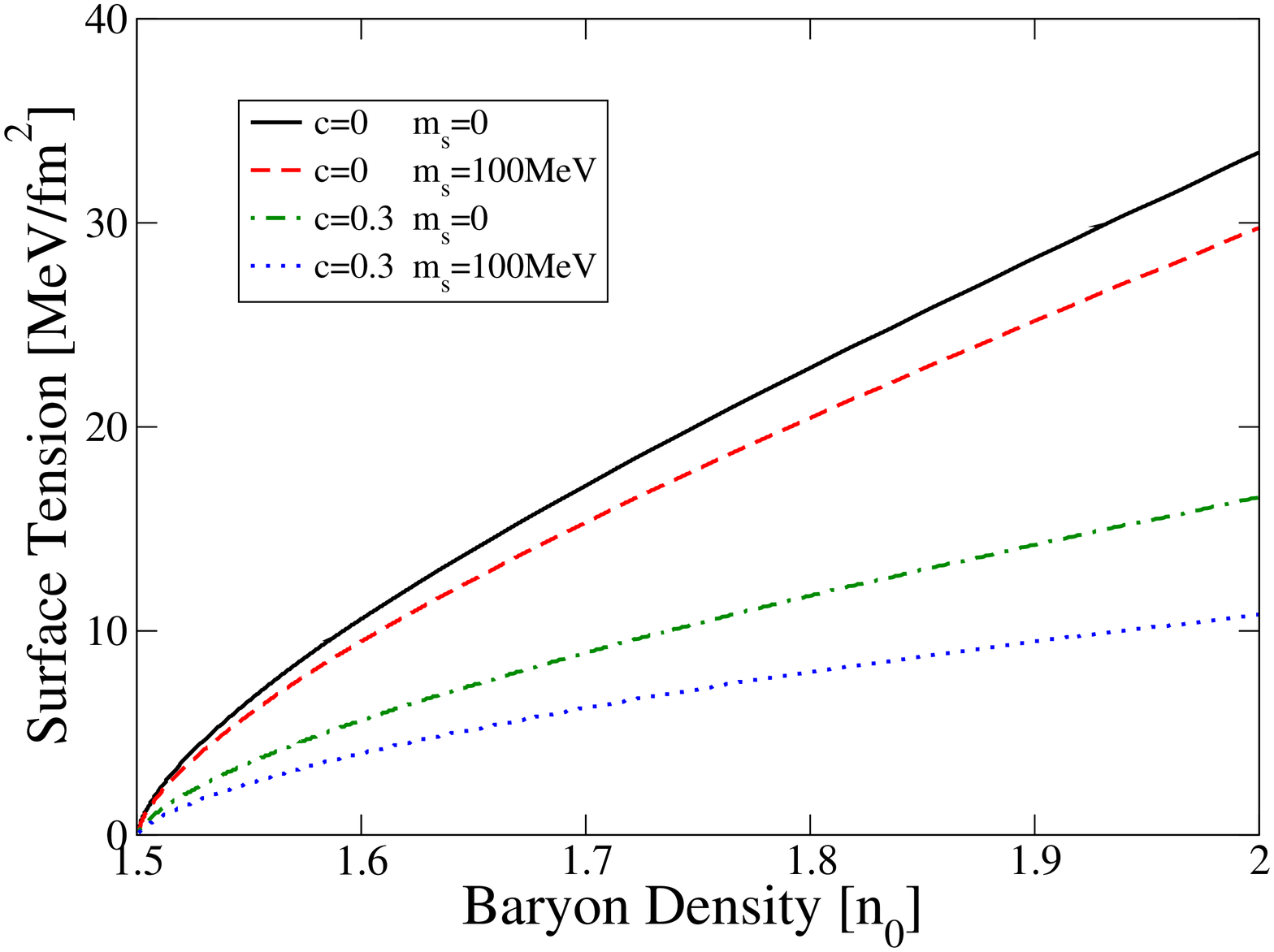}
\caption{Lines of constant nucleation time ($\tau=100~$ms) for 
$c=0$ and $c=0.3$, and for $m_s=0$ and $m_s=100$ MeV 
($n_c=1.5n_0$, $T=20~$MeV and $Y_L=0.4$).}
\end{center}
\end{figure}


\subsection{Stellar structure of selected equations of state}
In order to check if the equations of state we used are compatible
with observed pulsar data, we calculate their associated mass-radius
diagram using the TOV equations. In Table 1, we report the quark
model parameters, for strange matter, $Y_L =0.4$ and different choices 
of the critical density and the resulting maximum mass for cold and 
beta-stable stars. 

Notice that only by including the effect of QCD perturbative interactions 
is it possible to obtain masses for hybrid stars compatible with the recent
observation of PSR J1903+0327, $M=(1.671 \pm 0.008)M_{\odot}$
\cite{Freire:2009dr}.

\vspace*{0.5cm}
\begin{center}

 \begin{tabular}{c|c|c|c}
  $n_c/n_0$ & $c$ & $B^{1/4}$(MeV) & $M_{max}/M_\odot$ \crcr\hline
  1.2       & 0   & 159.22         &   1.60 \crcr
            & 0.3 & 144.65         &   1.90          \crcr\hline
  1.5       & 0   & 161.77         &   1.55  \crcr
            & 0.3 & 145.89         &   1.87          \crcr\hline
  2.0       & 0   & 166.64         &   1.48  \crcr
            & 0.3 & 147.56         &   1.83           \crcr

 \end{tabular}
\end{center}

\textbf{Table 1.} Maximum masses of cold deleptonized compact 
                  stars for some of the EoS used (corresponding  
                  to the cases $Y_L=0.4$ and $m_s=100$ MeV  
                  considered for nucleation).

\subsection{Quantum nucleation and spinodal instability}

There are other mechanisms which can compete with thermal nucleation
in driving the phase transition: quantum nucleation and spinodal decomposition.
Unfortunately, our treatment is blind to the spinodal instability: an effective potential 
is indeed needed to take it into account effectively (see Ref. \cite{Bessa:2008nw}).
In any case, this is a process that will be relevant only if the system is taken into 
values of density that are high enough to flatten out the activation barrier, so that 
there is no more extra cost to create a region of the true vacuum inside the initially 
homogeneous false vacuum configuration by thermal fluctuations. If that is possible, 
the phase conversion process will be rather explosive, a possibility we will consider 
in a forthcoming publication.

On the other hand, we can easily estimate the contribution of quantum nucleation 
by using the formalism considered in Ref. \cite{Iida:1997ay}, i.e. a WKB treatment of the 
tunneling through 
a barrier of an effective potential similar in form to Eq. (\ref{eq:DeltaF}).
We find that only for temperatures smaller than $\sim 5~$MeV
the quantum nucleation rate is comparable or larger than the thermal nucleation rate. 
Since we are considering supernova matter, with $T\gtrsim ~$10 MeV, thermal nucleation 
is by far the dominant mechanism for the production of the first drop of quark matter. 

\section{Conclusions}

We investigated the possibility of the formation of quark matter in
supernova matter, i.e. for temperatures of the order of a few tens of
MeV and in the presence of trapped neutrinos, assuming that the
corresponding critical density does not exceed $2 n_0$.  We argued
that thermal nucleation of quark phase droplets is eventually the
dominant mechanism for the formation of the new phase and that, due to
fluctuations in the number densities of quarks, the phase transition
involves directly the beta equilibrated quark phase. We have
calculated the nucleation rate for different values of the free
parameters. The surface tension, as expected, is the physical quantity
which mainly controls the nucleation process.

Among the different equations of state and conditions at bounce we
have tested, the choice $T=20$ MeV and $Y_L=0.4$ is the most likely to
occur. Within this choice of physical conditions, if the phase
transition involves only noninteracting up and down quarks, 
a value of $\sigma$
smaller than $\sim15~$MeV/fm$^2$ should be required for nucleation to be
efficient. Such a low value of $\sigma$ is compatible with lattice QCD
calculations, although they are obtained at large temperatures and small densities
\cite{deForcrand:2006ec}. On the other hand, phenomenological
estimates at large density and zero temperature indicate larger
values of $\sigma$ \cite{Voskresensky:2002hu}. Therefore, we consider
this scenario to be unlikely.

If strange quarks are produced during the phase transition, we
conclude that, if $\sigma$ is smaller than $\sim 10~$MeV/fm$^2$
(for $m_s=100$MeV and $c=0.3$), 
the nucleation time for the first drop of quark matter is sufficiently
small and the appearance of quark matter can indeed strongly affect
the supernova evolution as shown in \cite{Sagert:2008ka}. 
The cold hybrid stars obtained after deleptonization and cooling have, 
if perturbative QCD corrections are included in the equation of state,
maximum masses compatible with recent pulsar observations.

{\it Note added}- After finishing this work a paper has been published
which discusses similar issues \cite{Bombaci:2009jt}. 
In that work, the authors also conclude that nucleation is possible 
in protoneutron star matter. The main
difference between our work and Ref.~\cite{Bombaci:2009jt} is that in
the latter the thermal nucleation of quark matter is studied within
hot and deleptonized hadronic matter.

\acknowledgments

B.~W.~M. and E.~S.~F. thank CAPES, CNPq, FAPERJ, and FUJB/UFRJ for financial 
support. The work of G.~P. is supported by the Alliance Program of the 
Helmholtz Association (HA216/EMMI) and by the Deutsche
Forschungsgemeinschaft (DFG) under Grant No. PA1780/2-1.
J.~S.~B. is supported by the DFG through the Heidelberg Graduate School 
of Fundamental Physics. The authors also thank
the CompStar program of the European Science Foundation.
      

\bibliography{references}
\bibliographystyle{apsrev}
\end{document}